\documentclass[twocolumn,showpacs,preprintnumbers,amsmath,amssymb]{revtex4}

\usepackage{graphicx}
\usepackage{dcolumn}
\usepackage{bm}

\begin{document}

\preprint{UAB--FT--526}

\title{Scalar meson dynamics in Chiral Perturbation Theory}

\author{Albert Bramon}
\email{bramon@ifae.es}
\affiliation{Departament de F\'{\i}sica, Universitat Aut\`onoma de Barcelona, E-08193 Bellaterra (Barcelona), Spain}

\author{Rafel Escribano}
\email{Rafel.Escribano@ifae.es}
\affiliation{Grup de F\'{\i}sica Te\`orica and IFAE, Universitat Aut\`onoma de Barcelona, E-08193 Bellaterra (Barcelona), Spain}

\author{Jos\'e Lu\'{\i}s Lucio Mart\'{\i}nez}
\email{lucio@fisica.ugto.mx}
\affiliation{Instituto de F\'{\i}sica, Universidad de Guanajuato, Lomas del Bosque \# 103, Lomas del Campestre, 37150 Le\'on, Guanajuato, Mexico}


\begin{abstract}
A comparison of the linear sigma model (L$\sigma$M) and Chiral Perturbation Theory (ChPT) predictions
for pion and kaon dynamics is presented.
Lowest and next-to-leading order terms in the ChPT amplitudes are reproduced if one restricts to
scalar resonance exchange.
Some low energy constants of the order $p^4$ ChPT Lagrangian are fixed in terms of scalar meson masses.
Present values of these low energy constants are compatible with the L$\sigma$M dynamics.
We conclude that more accurate values would be most useful either to falsify the L$\sigma$M or
to show its capability to shed some light on the controversial scalar physics.
\end{abstract}

\maketitle

\section{\label{intro}Introduction}

Among the well know difficulties one has to face when dealing with low energy hadron physics,
those linked to the nature, properties and effects of scalar meson resonances are notoriously problematic
and continue to remain unsolved along the years.
In this sense, several interesting proposals have appeared concerning the constitution of these scalars
as multiquark states \cite{Jaffe:1976ig}, $K\bar K$ molecules \cite{Weinstein:gu} or
ordinary $q\bar q$ mesons, strongly distorted by unitary corrections \cite{Tornqvist:1995kr} or
without these drastic distortions.
But none of these proposals seems to be definitely preferable and discussions on the nature of the
scalar resonances are still open (see Ref.~\cite{Close:2002zu} for a recent discussion).
This controversial situation is probably related to the difficulties encountered when extracting the
main properties of the scalars from experimental data which are often affected by the opening of
two-pseudoscalar decay channels.
Indeed, a look at the current (or previous) PDG edition(s) \cite{Hagiwara:fs} shows a proliferation of
scalar states  above 1 GeV and conflictive or poor data ---usually affected by large uncertainties---
for their $f_{0}(980)$ and $a_{0}(980)$ partners with masses close to the $K\bar K$ threshold.
In particular, there is no consensus on the existence of the $\pi\pi$ scalar resonance $\sigma(600)$ and,
eventually, on the nature of this state.

At first sight, the situation concerning the $\sigma(600)$ has been considerably improved thanks to
two sets of recent experimental results.
The first set refers to radiative $\phi\to\pi^0\pi^0\gamma$ decays, as recently measured by the CMD-2 and
SND Collaborations at VEPP-2M \cite{Achasov:2000ym,Akhmetshin:1999di} and, with higher accuracy, by the
KLOE detector at DA$\Phi$NE \cite{Aloisio:2002bt}.
Another set of data comes from the $D_{s}^+\to\pi^-\pi^+\pi^+$ Dalitz-plot analysis performed by the
E791 Collaboration at Fermilab \cite{Aitala:2000xt}.
In all these experiments
\cite{Achasov:2000ym,Akhmetshin:1999di,Aloisio:2002bt,Aitala:2000xt}
one deals with a channel which is rich in strangeness, thus favoring the formation of intermediate
$K\bar K$ meson pairs.
The $f_{0}(980)$ is then eminently visible whereas the $\sigma(600)$ seems to be completely absent.
By contrast, the $\sigma(600)$ seems to play the dominant role in the strangeness-poor channel
$D^+\to\pi^-\pi^+\pi^+$ \cite{Aitala:2000xu}.
Although other interpretations are certainly possible and criticisms have been raised,
these results suggest that the $\sigma(600)$ couples strongly to pion pairs but not to kaon pairs.
In the $U(3)\times U(3)$ linear sigma model (L$\sigma$M), proposed many years ago by
Levy, Gasiorowicz, Schechter and coworkers \cite{LsM}, the $\sigma(600)$ couplings to pion and kaon pairs
are directly predicted to be proportional to $m_{\sigma}^2-m_{\pi}^2$ and $m_{\sigma}^2-m_K^2$, respectively.
The existence of such a low mass resonance of the L$\sigma$M, with $m_{\sigma}\simeq m_{K}$,
would thus immediately explain the above experimental findings
\cite{Achasov:2000ym,Akhmetshin:1999di,Aloisio:2002bt,Aitala:2000xt,Aitala:2000xu}.
This simple observation has renewed our interest on the L$\sigma$M.

Many theoretical discussions on scalar resonances have been published along the years based on the L$\sigma$M
\cite{Klabucar:2001gr,Ishida:1995xx,Shabalin:ey}
but we would like to concentrate on those recently appeared in
Refs.~\cite{Napsuciale:1998ip,Tornqvist:1999tn,'tHooft:1999jc,Black:2000qq}.
The main advantages of the L$\sigma$M are the central role played by the scalar states and the
high predictability of the model.
The drawback, however, is that the predictability becomes really effective only if a minimum of the
scalar meson properties are accurately known and can be used as a solid input to fix the model parameters.
This is clearly illustrated when comparing the conclusions of the four previously mentioned recent papers on the
L$\sigma$M \cite{Napsuciale:1998ip,Tornqvist:1999tn,'tHooft:1999jc,Black:2000qq}.
In spite of the fact that all of them are based on the same Lagrangian,
their predictions on the scalar properties are notoriously divergent \cite{Napsuciale:1998ip}.
For instance, the mass of the strange $\kappa$ and the $\sigma$ mesons are predicted to be, respectively,
around 900 and 375 MeV in Ref.~\cite{Napsuciale:1998ip}, 1120 and 650 MeV in Ref.~\cite{Tornqvist:1999tn},
and still higher values for these scalar masses follow from Ref.~\cite{'tHooft:1999jc}.
For many authors, the L$\sigma$M is thus a kind of ``toy model'' unable to account for the data even in a
first order approximation.

Unfortunately, the recently published data on radiative $\phi$ decays and three-pion decays of
$D$ and $D_{s}$ mesons do not represent a decisive improvement on our knowledge of the scalars.
Indeed, when the values of the various scalar couplings are extracted from the latter data,
contradictions with previous estimates appear, as briefly discussed in Ref.~\cite{Aitala:2000xt}.
Something similar happens with the interpretation of the dipion invariant mass spectrum in
$\phi\to\pi^0\pi^0\gamma$.
The data samples of the three experimental groups
\cite{Achasov:2000ym,Akhmetshin:1999di,Aloisio:2002bt}
are quite compatible and their analyses are similarly based on the same kaon-loop mechanism
\cite{Close:ay,Achasov:1987ts,LucioMartinez:uw}.
According to this common mechanism, $K^+K^-$ pairs copiously produced in $\phi$ decays rescatter into
$\pi^0\pi^0$ through a scalar dominated $s$-channel after the emission of a photon.
But the $\sigma(600)$ contribution is simply not considered or found to have negligible effects in the
analyses of Refs.~\cite{Achasov:2000ym,Akhmetshin:1999di} (see also \cite{Bramon:2002iw}),
whereas it plays a major role according to the analysis of Ref.~\cite{Aloisio:2002bt}.
Since this unpleasant situation concerning scalar states has been lasting for many years,
we propose to adopt a different strategy.
Namely, to complement the L$\sigma$M with the guidance of the solid theoretical framework of
Chiral Perturbation Theory (ChPT),
rather than with the direct use of poorly known and disputable experimental data on scalar resonances.

Nowadays ChPT is considered to be the appropriate framework to discuss hadronic interactions at low energy
\cite{Gasser:1983yg}.
Leading role in ChPT is reserved to the octet of pseudoscalar mesons, hereafter denoted $P_{8}$,
entering as external lines in the various diagrams and also as internal lines in chiral loops.
The effects of meson resonances
---including those generated by the exchange of low mass scalar states---
are assumed to manifest in ChPT through the low energy constants in the various counterterms needed to
cancel the pseudoscalar loop divergencies.
The values of these low energy constants are assumed to be saturated by resonance exchange effects in
their corresponding channels \cite{Gasser:1983yg,Donoghue:ed,Ecker:1988te,Bijnens:1988kx}.
This saturation and other relationships between the L$\sigma$M and ChPT have been discussed by
several authors at different levels with somewhat conflicting conclusions.
In Refs.~\cite{Donoghue:ed,Donoghue:dd}, for instance,
the inability of scalar resonance exchange as dictated by the L$\sigma$M to account for the
ten next-to-leading terms of the ChPT Lagrangian is seen as a proof against L$\sigma$M dynamics.
Less explicitly, this criticism is similarly mentioned in the recent and comprehensive treatment of the
L$\sigma$M in Ref.~\cite{Tornqvist:1999tn}.
We certainly agree on the inability of scalar resonances \emph{alone} to saturate
\emph{all} ten low energy constants of the ChPT Lagrangian, $L_{1,\ldots,10}$,
but we still believe that they do saturate the part of these constants corresponding to scalar exchange.
In this sense, we adhere to the proposal by Ecker \textit{et al.}~\cite{Ecker:1988te},
or the analysis in Ref.~\cite{Bramon:1994bw}, and try to take advantage of the present knowledge of the
appropriate terms of the ChPT Lagrangian to shed some light into the confuse and controversial sector of
scalar mesons.
As we will see, the information one can extract is not detailed but provides an average behavior over the
whole scalar multiplet(s).
But, because of this, it is also free from details concerning the opening of individual channels at
different thresholds and thus avoids one of the major problems encountered in scalar data analyses.
To this aim, we have to start discussing the compatibility between the L$\sigma$M and ChPT.
These two approaches are closely related and valuable information on the scalar meson dynamics is contained in
various terms of the ChPT Lagrangian.

\section{L$\sigma$M and ChPT Lagrangians}
The ChPT Lagrangian is written in terms of the octet of pseudoscalar (Goldstone) mesons, $P_{8}$,
and the electroweak gauge bosons, which need not to be considered for our present purposes.
In this case and at lowest order, it contains a symmetrical kinetic term and a mass term which breaks the
$SU(3)$ symmetry: 
\begin{equation}
\label{L2}
{\cal L}_{2}=\frac{f^2}{4}\langle\partial_{\mu}U^\dagger\partial^{\mu}U\rangle
            +\frac{f^2}{4}2B_{0}\langle{\cal M}(U+U^\dagger)\rangle\ , 
\end{equation}
where $U=\exp\left(i\sqrt{2}\Phi/f\right)$, $\Phi\equiv\frac{1}{\sqrt{2}}\sum_{i=1}^8\lambda_{i}\phi_{i}^{(P)}$,
$\phi_{i}^{(P)}$ are the eight pseudoscalar fields and $\lambda_{i}$ the Gell-Mann matrices.  
Besides the pseudoscalar decay constant $f=f_{P_{8}}$, which at lowest order is common to all the octet members,
and the different quark masses appearing via ${\cal M}=\mbox{diag}(m_{u}, m_{d}, m_{s})$,
other ten low energy constants, $L_{i}$, are required to express the next order piece of the ChPT Lagrangian 
\begin{equation}
\label{L4}
{\cal L}_{4}=\sum_{i=1}^{10}L_{i}{\cal L}_{4}^{(i)}\ ,  
\end{equation}
with ${\cal L}_{4}^{(i)}$ terms explicitly given in Ref.~\cite{Gasser:1983yg} and briefly commented below.

Six of these constants, $L_{1,3,4,5,6,8}$, are known to contain the effects of scalar resonance exchange
\cite{Ecker:1988te} and are the relevant ones in our present discussion.
$L_{5,8}$ are found to be saturated by the exchange of the scalar octet alone,
$S_{8}$, while $L_{3}$ contains contributions from both the scalar and vector meson octets,
\textit{i.e.}~$L_{5,8}=L_{5,8}^{S_{8}}$ and $L_{3}=L_{3}^{S_{8}}+L_{3}^{V}$.
Something similar happens to the other three relevant low energy constants.
$L_{4,6}$ are saturated by the exchange of scalar resonances, while $L_{1}$ contains contributions from
both the scalar and vector meson multiplets, $L_{4,6}=L_{4,6}^{S}$ and $L_{1}=L_{1}^{S}+L_{1}^{V}$.
Remark however that the scalar effects in $L_{1,4,6}$ involve the whole scalar \emph{nonet},
$S=S_{8}+S_{0}$, and turn out to be proportional to the octet-singlet mass difference.
Since this mass difference vanishes in the large $N_{c}$ limit, one usually assumes following
Ecker \textit{et al.}~\cite{Ecker:1988te} that these three low energy constants receive no
important contributions from the scalar resonances,
\textit{i.e.}~$L_{1,4,6}^{S_{8}}+L_{1,4,6}^{S_{0}}\equiv L_{1,4,6}^{S}\simeq 0$,
a rather drastic approximation to be improved by our present L$\sigma$M analysis.
Note that the characteristics of the six terms multiplying the low energy constants are rather different:
$L_{1,3}$ appear in $SU(3)$-symmetric terms consisting of four derivative factors,
$L_{4,5}$ appear in terms with two derivative and one $SU(3)$-breaking (${\cal M}$) factors,
and the $L_{6,8}$ terms are non-derivative and contain two $SU(3)$-breaking (${\cal M}^2$) factors.

The L$\sigma$M Lagrangian contains also $SU(3)$-symmetric and $SU(3)$-breaking terms but,
apart from this feature, it is different from the ChPT Lagrangian.
Indeed, although the L$\sigma$M Lagrangian is also written in terms of the pseudoscalar octet $P_{8}$
it explicitly contains the pseudoscalar singlet field, $P_{0}=\eta_{0}$,
and the whole nonet, $S=S_{8}+S_{0}$, of scalar fields.
All these eighteen fields, $\phi_{i}^{(P)}$ and $\phi_{i}^{(S)}$, with $i=0, 1,\ldots, 8$,
are organized linearly in the meson matrix 
\begin{equation}
\label{M}
M\equiv S+iP\ ,\quad S,P=\frac{1}{\sqrt{2}}\sum_{i=0}^{8}\lambda_{i}\phi_{i}^{(S,P)}\ .
\end{equation}
The presence of the $\eta_{0}$ singlet induces $\eta$-$\eta^\prime$ mixing effects
which are well defined in the L$\sigma$M but not in ChPT,
where they proceed through an adjustable $L_{7}$ term in the Lagrangian.
Moreover, in the L$\sigma$M there are no derivative terms
---apart from those corresponding to the kinetic energy given by the trace of
$\frac{1}{\sqrt{2}}\partial_{\mu}M\partial^{\mu}M^\dagger$.
The L$\sigma$M terms accounting for the interactions are non-derivative and their expression
can be found in the recent Refs.~\cite{Napsuciale:1998ip,Tornqvist:1999tn,'tHooft:1999jc,Black:2000qq}.
They contain the trace of several $M$-factors, such as $M M^\dagger$ and $M M^\dagger M M^\dagger$,
the squared trace of $M M^\dagger$ or the trace of an odd number of $M$ matrices in the
two symmetry breaking terms.
This in sharp contrast with ChPT, where derivative couplings are required by the Goldstone boson nature
of the pseudoscalars.

Thus, even if the $\eta_{0}$ and $S=S_{8}+S_{0}$ fields are integrated out,
the structure of the L$\sigma$M and ChPT Lagrangians remains completely different.
For this reason, instead of matching at the Lagrangian level,
we have to compare the predictions of both approaches for the various physical properties
(masses and decay constants) and scattering amplitudes for the pseudoscalar octet $P_{8}$.
In this paper, we show that L$\sigma$M and ChPT expressions for $f_{\pi,K}$, $m_{\pi,K}^2$ and for the
$\pi\pi\to\pi\pi$, $\pi K\to\pi K$ and $K\bar K\to K\bar K$ amplitudes consistently predict the values of
$L_{1,3,4,5,6,8}$ in terms of the scalar resonance masses and mixing angle.
Due to the high predictability of the L$\sigma$M, these predictions are overconstrained and there is thus
no need to consider the eighth pseudoscalar state, $\eta_{8}$, which complicates the analysis by mixing
with the singlet $\eta_{0}$ treated so differently in the L$\sigma$M and in ChPT (see below).

\section{L$\sigma$M results}
From any of the recent analyses on the L$\sigma$M in
Refs.~\cite{Napsuciale:1998ip,Tornqvist:1999tn,'tHooft:1999jc,Black:2000qq} 
one can easily deduce the following relation
\begin{equation}
\label{fKoverfpiLsM}
\frac{f_{K}}{f_{\pi}}=\frac{M_{\kappa}^2-m_{\pi}^2}{M_{\kappa}^2-m_{K}^2}
                     =1+\frac{m_{K}^2-m_{\pi}^2}{M_{\kappa}^2}+\cdots\ ,
\end{equation}
where $M_{\kappa}^2$ is the squared mass of the strange scalar resonance,
and $m_{\pi,K}^2$ and $f_{\pi,K}$ are the squared masses and decay constants of the
pion and kaon isomultiplets.
The nine scalar mesons are assumed to be much heavier than pions and kaons,
$M_{S_{8},S_{0}}^2\gg m_{\pi,K}^2$, thus allowing for a series expansion.
The ellipsis in Eq.~(\ref{fKoverfpiLsM}) and the following L$\sigma$M expressions,
refer to terms of order $1/M_{S_{8},S_{0}}^4$ or higher, that will be systematically neglected.
We work in the good isospin limit ($m_{u}=m_{d}$) and present the results in terms of physical quantities
thus avoiding the use of different notations introduced in 
Refs.~\cite{Napsuciale:1998ip,Tornqvist:1999tn,'tHooft:1999jc,Black:2000qq}.
In particular, we use $f_{\pi}=92.4$ MeV.

The L$\sigma$M expressions for the scattering amplitudes are somewhat more involved.
As well known, the three isospin amplitudes, $I=0,1,2$, governing $\pi\pi\to\pi\pi$ scattering can be
expressed in terms of a single amplitude $T(\pi^+\pi^-\to\pi^0\pi^0)$.
For the latter one easily obtains
\begin{widetext}
\begin{equation}
\label{TpipipipiLsM}
T_{\pi^+\pi^-\to\pi^0\pi^0}^{\mbox{\scriptsize L$\sigma$M}}=
\frac{s-m_{\pi}^2}{f_{\pi}^2}\left[1+(s-m_{\pi}^2)\left(\frac{\cos^2\phi_{S}}{M_{\sigma}^2}+
                                   \frac{\sin^2\phi_{S}}{M_{f_{0}}^2}\right)+\cdots\right]\ ,
\end{equation}
\end{widetext}
where $M_{\sigma,f_{0}}^2$ are the squared masses of the two physical $I=0$ scalar states and
$\phi_{S}$ is their mixing angle in the non-strange--strange quark basis $(\sigma_{\rm NS}, \sigma_{S})$
\begin{eqnarray}
\sigma=\cos\phi_{S}\,\sigma_{\rm NS}-\sin\phi_{S}\,\sigma_{S}\ ,\nonumber\\
f_{0}=\sin\phi_{S}\,\sigma_{\rm NS}+\cos\phi_{S}\,\sigma_{S}\ .
\label{quarkbasis}
\end{eqnarray}

For the $K\pi\to K\pi$ channel there are two isospin amplitudes but that for $I=1/2$ can be deduced by
crossing symmetry from the $I=3/2$ amplitude.
For the latter amplitude the L$\sigma$M predicts
\begin{widetext}
\begin{eqnarray}
T_{K^+\pi^+\to K^+\pi^+}^{\mbox{\scriptsize L$\sigma$M}}&=&
\frac{1}{2f_{\pi}f_{K}}\left\{t+u-m_{\pi}^2-m_{K}^2+(t-m_{\pi}^2)(t-m_{K}^2)\right.\nonumber\\
&&\times\left[
\frac{{\rm c}\phi_{S}({\rm c}\phi_{S}-\sqrt{2}\,{\rm s}\phi_{S})}{M_{\sigma}^2}+
\frac{{\rm s}\phi_{S}({\rm s}\phi_{S}+\sqrt{2}\,{\rm c}\phi_{S})}{M_{f_{0}}^2}+
\frac{(u-m_{\pi}^2)(u-m_{K}^2)}{M_{\kappa}^2}+\cdots\right\}\ ,
\label{TKpiKpiLsM}
\end{eqnarray}
\end{widetext}
where $({\rm c}\phi_S, {\rm s}\phi_S)\equiv (\cos\phi_S, \sin\phi_S)$.
Note that at this level we distinguish between $f_{\pi}$ and $f_{K}$ and among the various scalar masses.
Our L$\sigma$M expressions are thus exact except for terms of order $1/M_{S_{8},S_{0}}^4$ or higher.

The two isospin amplitudes for $K\bar K\to K\bar K$ scattering are independent and can be deduced from
\begin{widetext}
\begin{eqnarray}
T_{K^+K^-\to K^+K^-}^{\mbox{\scriptsize L$\sigma$M}}&=&
\frac{1}{f_{K}^2}\left\{s+t-2m_{K}^2+\frac{(s-m_{K}^2)^2+(t-m_{K}^2)^2}{4}\right.\nonumber\\
&&\left.\times\left[
\frac{({\rm c}\phi_{S}-\sqrt{2}\,{\rm s}\phi_{S})^2}{M_{\sigma}^2}+
\frac{({\rm s}\phi_{S}+\sqrt{2}\,{\rm c}\phi_{S})^2}{M_{f_{0}}^2}+
\frac{1}{M_{a_{0}}^2}\right]+\cdots\right\}\ ,
\label{TKKKKchLsM}
\end{eqnarray}
and
\begin{eqnarray}
T_{K^+K^-\to K^0\bar K^0}^{\mbox{\scriptsize L$\sigma$M}}&=&
\frac{1}{2f_{K}^2}\left\{s+t-2m_{K}^2+
\frac{(t-m_{K}^2)^2}{M_{a_{0}}^2}+\frac{(s-m_{K}^2)^2}{2}\right.\nonumber\\
&&\left.\times\left[
\frac{({\rm c}\phi_{S}-\sqrt{2}\,{\rm s}\phi_{S})^2}{M_{\sigma}^2}+
\frac{({\rm s}\phi_{S}+\sqrt{2}\,{\rm c}\phi_{S})^2}{M_{f_{0}}^2}-
\frac{1}{M_{a_{0}}^2}\right]+\cdots\right\}\ .
\label{TKKKKneuLsM}
\end{eqnarray}
\end{widetext}

This completes the L$\sigma$M results we need to consider.
Note that all these amplitudes have the required Adler zeroes and vanish when any of the
pseudoscalar four-momenta is sent to zero.
Thanks to this feature, our results can be expressed in a compact form and in terms of physical quantities.

\section{ChPT results}
Following the same order as in the preceding section below we list the ChPT predictions.
They include the leading (lowest order) term and the contributions from the six counterterms,
$L_{1,3,4,5,6,8}$, affected by the scalar meson exchange we are considering.
Needless to say, these ChPT amplitudes at the next-to-leading order should be completed by
one-loop contributions and by those coming from non-scalar exchange.
For the ratio of decay constants one has the well known ChPT result
\begin{equation}
\label{fKoverfpiChPT}
\frac{f_{K}}{f_{\pi}}=1+\frac{4L_{5}}{f_{\pi}f_{K}}(m_{K}^2-m_{\pi}^2)+\cdots\ ,
\end{equation}
where the ellipsis stands for the non-scalar exchange effects that we are systematically neglecting.

For the various pion and kaon scattering amplitudes we recover the results of Refs.~\cite{Gasser:1983yg}
and \cite{Bernard:1990kw,Dobado:1992ha,Guerrero:1998ei,GomezNicola:2001as}:
\begin{widetext}
\begin{equation}
T_{\pi^+\pi^-\to\pi^0\pi^0}^{\mbox{\scriptsize ChPT}}=
\frac{s-m_{\pi}^2}{f_{\pi}^2}
+\frac{4}{f_{\pi}^4}\left[(2L_{1}^S+L_{3}^S)(s-2m_{\pi}^2)^2
+2(2L_{4}+L_{5})m_{\pi}^2(s-2m_{\pi}^2)+4(2L_{6}+L_{8})m_{\pi}^4+\cdots\right]\ ,
\label{TpipipipiChPT}
\end{equation}
\begin{eqnarray}
T_{K^+\pi^+\to K^+\pi^+}^{\mbox{\scriptsize ChPT}}&=&
\frac{t+u-m_{\pi}^2-m_{K}^2}{2f_{\pi}f_{K}}+\frac{4}{f_{\pi}^2 f_{K}^2}
\times\left\{(4L_{1}^S+L_{3}^S)(t-2m_{\pi}^2)(t-2m_{K}^2)+L_{3}^S(u-m_{\pi}^2-m_{K}^2)^2\right.\nonumber\\
&&+4L_{4}\left[t(m_{\pi}^2+m_{K}^2)-4m_{\pi}^2 m_{K}^2\right]
+L_{5}\left[(m_{\pi}^2+m_{K}^2)(t+u-m_{\pi}^2-m_{K}^2)-4m_{\pi}^2 m_{K}^2\right].\nonumber\\
&&\left.+8(2L_{6}+L_{8})m_{\pi}^2 m_{K}^2+\cdots\right\}\ ,
\label{TKpiKpiChPT}
\end{eqnarray}
\begin{eqnarray}
T_{K^+K^-\to K^+K^-}^{\mbox{\scriptsize ChPT}}&=&
\frac{s+t-2m_{K}^2}{f_{K}^2}+\frac{4}{f_{K}^4}
\times\left\{(2L_{1}^S+L_{3}^S)\left[(s-2m_{K}^2)^2+(t-2m_{K}^2)^2\right]\right.\nonumber\\
&&\left.-2(2L_{4}+L_{5})u m_{K}^2+8(2L_{6}+L_{8})m_{K}^4+\cdots\right\}\ ,
\label{TKKKKchChPT}
\end{eqnarray}
and
\begin{eqnarray}
T_{K^+K^-\to K^0\bar K^0}^{\mbox{\scriptsize ChPT}}&=&
\frac{s+t-2m_{K}^2}{2f_{K}^2}+\frac{2}{f_{K}^4}
\times\left\{(4L_{1}^S+L_{3}^S)(s-2m_{K}^2)^2+L_{3}^S(t-2m_{K}^2)^2\right.\nonumber\\
&&\left.+2(4L_{4}+L_{5})s m_{K}^2+2L_{5}t m_{K}^2-8(2L_{4}+L_{5})m_{K}^4
+8(2L_{6}+L_{8})m_{K}^4+\cdots\right\}\ .
\label{TKKKKneuChPT}
\end{eqnarray}
\end{widetext}
Note that the kinematical constraint $s+t+u=\sum m_{i}^2$,
where the sum extends to the four pseudoscalar masses involved in each process,
has been used to express the amplitudes (\ref{TpipipipiChPT})--(\ref{TKKKKneuChPT}) in terms of
the same kinematical variables as in the corresponding L$\sigma$M expressions
(\ref{TpipipipiLsM})--(\ref{TKKKKneuLsM}).

\section{Confronting the L$\sigma$M with ChPT}
We are now ready to compare the results of the previous two sections.
The simplest case corresponds to Eqs.~(\ref{fKoverfpiLsM}) and (\ref{fKoverfpiChPT}) from which
one immediately obtains
\begin{equation}
\label{L5}
L_{5}=\frac{f_{\pi}f_{K}}{4M_{\kappa}^2}\ .
\end{equation}

To identify the remaining low energy constants we consider the scattering amplitudes.
The leading terms, which contain no scalar masses or $L_{i}$'s constants,
are seen to be equivalent in both approaches as required by chiral symmetry.
Similarly, next-to-leading terms have also the same structure showing thus the compatibility between
L$\sigma$M and ChPT, even if $SU(3)$-breaking effects are retained.
From a direct comparison of the $\pi^+\pi^-\to\pi^0\pi^0$ amplitudes one finds
\begin{widetext}
\begin{equation}
\label{comppipi}
2L_{1}^S+L_{3}^S=2L_{4}+L_{5}=4(2L_{6}+L_{8})=
\frac{f_{\pi}^2}{4}\left(\frac{\cos^2\phi_{S}}{M_{\sigma}^2}+\frac{\sin^2\phi_{S}}{M_{f_{0}}^2}\right)\ ,
\end{equation}
\end{widetext}
while from $K^+K^-\to K^+K^-$ scattering one gets
\begin{widetext}
\begin{equation}
2L_{1}^S+L_{3}^S=2L_{4}+L_{5}=4(2L_{6}+L_{8})=
\frac{f_{K}^2}{16}\left[
\frac{({\rm c}\phi_{S}-\sqrt{2}\,{\rm s}\phi_{S})^2}{M_{\sigma}^2}+
\frac{({\rm s}\phi_{S}+\sqrt{2}\,{\rm c}\phi_{S})^2}{M_{f_{0}}^2}+
\frac{1}{M_{a_{0}}^2}\right]\ .
\label{compKKch}
\end{equation}
\end{widetext}

The comparison of $K^+\pi^+\to K^+\pi^+$ and $K^+K^-\to K^0\bar K^0$ scattering amplitudes is richer and
allows to fix some of the $L_{i}$'s individually.
From $K\pi$ scattering amplitude one obtains
\begin{widetext}
\begin{equation}
\label{compKpiL35}
L_{3}^S=L_{5}=\frac{f_{\pi}f_{K}}{4M_{\kappa}^2}\ ,
\end{equation}
\begin{equation}
\label{compKpiL14}
L_{1}^S=L_{4}=\frac{f_{\pi}f_{K}}{16}\left[
\frac{{\rm c}\phi_{S}({\rm c}\phi_{S}-\sqrt{2}\,{\rm s}\phi_{S})}{M_{\sigma}^2}+
\frac{{\rm s}\phi_{S}({\rm s}\phi_{S}+\sqrt{2}\,{\rm c}\phi_{S})}{M_{f_{0}}^2}-
\frac{1}{M_{\kappa}^2}\right]\ ,
\end{equation}
\begin{equation}
\label{compKpi}
2L_{6}+L_{8}=\frac{f_{\pi}f_{K}}{32}\left[
\frac{{\rm c}\phi_{S}({\rm c}\phi_{S}-\sqrt{2}\,{\rm s}\phi_{S})}{M_{\sigma}^2}+
\frac{{\rm s}\phi_{S}({\rm s}\phi_{S}+\sqrt{2}\,{\rm c}\phi_{S})}{M_{f_{0}}^2}+
\frac{1}{M_{\kappa}^2}
\right]\ ,
\end{equation}
\end{widetext}
while $K\bar K$ scattering requires
\begin{widetext}
\begin{equation}
\label{compKKneuL35}
L_{3}^S=L_{5}=\frac{f_{K}^2}{4M_{a_{0}}^2}\ ,
\end{equation}
\begin{equation}
\label{compKKneuL14}
L_{1}^S=L_{4}=\frac{f_{K}^2}{32}\left[
\frac{({\rm c}\phi_{S}-\sqrt{2}\,{\rm s}\phi_{S})^2}{M_{\sigma}^2}+
\frac{({\rm s}\phi_{S}+\sqrt{2}\,{\rm c}\phi_{S})^2}{M_{f_{0}}^2}-
\frac{3}{M_{a_{0}}^2}\right]\ ,
\end{equation}
\begin{equation}
\label{compKKneu}
2L_{6}+L_{8}=\frac{f_{K}^2}{64}\left[
\frac{{\rm c}\phi_{S}({\rm c}\phi_{S}-\sqrt{2}\,{\rm s}\phi_{S})}{M_{\sigma}^2}+
\frac{{\rm s}\phi_{S}({\rm s}\phi_{S}+\sqrt{2}\,{\rm c}\phi_{S})}{M_{f_{0}}^2}+
\frac{1}{M_{a_{0}}^2}\right]\ .
\end{equation}
\end{widetext}
$L_{6}$ and $L_{8}$ cannot be fixed individually but only in the particular combination $2L_{6}+L_{8}$.

Remember that according to Eq.~(\ref{fKoverfpiLsM}) corrections to $f_{K}/f_{\pi}$ are of order $1/M_{S}^2$.
Therefore, in comparing the leading order results of ChPT and the L$\sigma$M we have to maintain the
flavor dependence of the pseudoscalar decay constants, \textit{i.e.}~$f_{K}\neq f_{\pi}$.
On the other hand, when comparing the predictions for the $1/M_{S}^2$ corrections we can use $f_{K}=f_{\pi}$
since we are not interested in the $1/M_{S}^4$ terms.
Moreover, we neglect the flavor dependence of the scalar octet $(M_{a_{0}}^2=M_{\kappa}^2\equiv M_{f_{8}}^2)$.
In this $SU(3)$ limit, the scalar octet decouples from the singlet and the scalar mixing angle in the
non-strange--strange quark basis is fixed to $\sin\phi_{S}=-1/\sqrt{3}$.
In this case, the expressions in Eqs.~(\ref{L5})--(\ref{compKKneu}), which are equal up to
$SU(3)$-breaking corrections, consistently give
\begin{widetext}
\begin{equation}
L_{3}^S=L_{5}=\frac{f_{\pi}^2}{4M_{S_{8}}^2}\ ,\quad
L_{1}^S=L_{4}=\frac{f_{\pi}^2}{12}\left(\frac{1}{M_{S_{0}}^2}-\frac{1}{M_{S_{8}}^2}\right)\ ,\quad
2L_{6}+L_{8}=\frac{f_{\pi}^2}{48}\left(\frac{2}{M_{S_{0}}^2}+\frac{1}{M_{S_{8}}^2}\right)\ ,
\label{LiSU3limit}
\end{equation}
\end{widetext}
where $M_{S_{0}}$ is the mass of the scalar singlet and $M_{S_{8}}$ the mean mass of the octet,
$M_{S_{8}}^2 \equiv\frac{1}{8}\left(3M_{a_{0}}^2+4M_{\kappa}^2+M_{f_{8}}^2\right)$.
These Eqs.~(\ref{LiSU3limit}) are our main results.

\section{Comparison and conclusions}
As previously mentioned, the role of scalar resonances in ChPT has been analyzed by other authors.
Only in a few cases, however, the precise L$\sigma$M dynamics is invoked and the results are then
presented without many details.
For instance, the expression for $L_{5}$ in Eq.~(\ref{LiSU3limit}) is not new,
it reproduces Leutwyler's result in Ref.~\cite{Leutwyler:1997yr}.
Another analysis can be found in the well-known book by Donoghue \textit{et al.}~\cite{Donoghue:dd},
where the following relations are presented,
\begin{equation}
\label{LisDonoghue}
2L_{1}^S+L_{3}^S=2L_{4}+L_{5}=4(2L_{6}+L_{8})=\frac{f_{\pi}^2}{4M_{S}^2}\ .
\end{equation}
Here $M_{S}^2$ stands for a generic scalar mass and agreement with our results is achieved once
the differences among all the scalar masses are ignored.

Scalar resonance contributions to the ChPT low energy constants, $L_{i}$,
can be also computed from suitable Lagrangians including the coupling of scalars to two pseudoscalars
\cite{Ecker:1988te,Bernard:1991zc,Amoros:2000mc}.
In this approach one has
\begin{eqnarray}
&&L_{1}^{S_{8}+S_{0}}=\frac{\tilde c_{d}^2}{2M_{S_{0}}^2}-\frac{c_{d}^2}{6M_{S_{8}}^2}\ ,\quad
L_{3}^{S_{8}}=\frac{c_{d}^2}{2M_{S_{8}}^2}\ ,\nonumber\\
&&L_{4}^{S_{8}+S_{0}}=\frac{\tilde c_{d}\tilde c_{m}}{M_{S_{0}}^2}-\frac{c_{d}c_{m}}{3M_{S_{8}}^2}\ ,\quad
L_{5}^{S_{8}}=\frac{c_{d}c_{m}}{M_{S_{8}}^2}\ ,\nonumber\\
&&2L_{6}^{S_{8}+S_{0}}+L_{8}^{S_{8}}=\frac{\tilde c_{m}^2}{M_{S_{0}}^2}+\frac{\tilde c_{m}^2}{6M_{S_{8}}^2}\ ,
\label{LisEcker}
\end{eqnarray}
where $c_{d}(\tilde c_{d})$ and $c_{m}(\tilde c_{m})$ are the constants of the derivative and
massive terms coupling the scalar octet(singlet) to two pseudoscalars.
Comparing these predictions (\ref{LisEcker}) with the L$\sigma$M results in the $SU(3)$ limit
(\ref{LiSU3limit}), one gets
\begin{eqnarray}
&&|c_{d}|=\sqrt{3}\,|\tilde c_{d}|=\frac{f_{\pi}}{\sqrt{2}}\ ,\nonumber\\
&&|c_{m}|=\sqrt{3}\,|\tilde c_{m}|=\frac{f_{\pi}}{2\sqrt{2}}\ ,\quad
  c_{d}c_{m}, \tilde c_{d}\tilde c_{m}>0\ ,
\label{cs}
\end{eqnarray}
without invoking large--$N_{c}$ arguments.
For $N_{c}\to\infty$, $M_{S_{8}}=M_{S_{0}}\equiv M_{S}$ and the L$\sigma$M predictions in
Eq.~(\ref{LiSU3limit}) reduce to
\begin{eqnarray}
&&L_{3}^S=L_{5}=\frac{f_{\pi}^2}{4M_{S}^2}\ ,\quad L_{1}^S=L_{4}=0\ ,\nonumber\\
&&2L_{6}+L_{8}=\frac{f_{\pi}^2}{16M_{S}^2}\ ,
\label{LislargeNc}
\end{eqnarray}
in agreement with the analyses in Refs.~\cite{Ecker:1988te,Leutwyler:1997yr}.

Numerical estimates of the L$\sigma$M contribution to the low energy constants could be obtained from
Eqs.~(\ref{LiSU3limit}), once $f_\pi$ and the values for $M_{S_0}$ and $M_{S_8}$ are given.
Unfortunately the scalar meson masses are poorly known and their corresponding L$\sigma$M predictions
depend strongly on the input chosen \cite{Napsuciale:1998ip,Tornqvist:1999tn,'tHooft:1999jc,Black:2000qq}.
For instance, taking $m_{\pi,K}^2$, $f_{\pi,K}$ and $m_{\eta}^2+m_{\eta^\prime}^2$ as input values,
the L$\sigma$M Lagrangian predicts for the mean octet mass
\begin{equation}
\label{predmass8LsM}
M_{S_8}=(1.1\pm 0.1)\ \mbox{GeV}\ .
\end{equation}
On the other hand $M_{S_0}$ cannot be fixed without additional inputs from the scalar sector.
With a somewhat enlarged error to account for the poor knowledge on this scalar sector, one expects
\cite{Napsuciale:1998ip,Tornqvist:1999tn,'tHooft:1999jc,Black:2000qq}
\begin{equation}
\label{predmass0LsM}
M_{S_0}=(0.8\pm 0.2)\ \mbox{GeV}\ .
\end{equation}
From these values and Eqs.~(\ref{LiSU3limit}) one gets
\begin{eqnarray}
&&L_{3}^S=L_{5}=(1.8\pm 0.3)\ 10^{-3}\ ,\nonumber\\
&&L_{1}^S=L_{4}=(0.5\pm 0.4)\ 10^{-3}\ ,\nonumber\\
&&2L_{6}+L_{8}=(0.6\pm 0.2)\ 10^{-3}\ ,
\label{predLisLsM}
\end{eqnarray}
to be compared with the following independent phenomenological estimates.

In Ref.~\cite{Amoros:2000mc} an estimate is obtained assuming $L_{1}^S=L_{4}=L_{6}=0$.
In this detailed study one finds
\begin{eqnarray}
&&L_{5}=(0.91\pm 0.15)\ 10^{-3}\ ,\nonumber\\
&&L_{8}=(0.62\pm 0.20)\ 10^{-3}\ .
\label{LisBijnens}
\end{eqnarray}
A similar analysis performed in Ref.~\cite{Leutwyler:1997yr} leads to
\begin{equation}
\label{LisLeutwyler}
L_{5}\simeq 2.2\ 10^{-3}\ ,\quad L_{8}\simeq 1.0\ 10^{-3}\ .
\end{equation}
Other authors have not imposed the large $N_c$ prediction $L_{1}^S=L_{4}=L_{6}=0$.
For instance in Ref.~\cite{Ecker:1988te}, by now a classical paper, Ecker \textit{et al.}~obtained a
model independent determination of the $L$'s:
\begin{eqnarray}
&&L_{4}=(-0.3\pm 0.5)\ 10^{-3}\ ,\nonumber\\
&&L_{6}=(-0.2\pm 0.3)\ 10^{-3}\ ,\nonumber\\
&&L_{5}=(1.4\pm 0.5)\ 10^{-3}\ ,\nonumber\\
&&L_{8}=(0.9\pm 0.3)\ 10^{-3}\ .
\label{LisEcker2}
\end{eqnarray}
In the context of the Inverse Amplitude Method of Ref.~\cite{GomezNicola:2001as},
these latter results can be improved to
$L_{4}=(-0.36\pm 0.17)\ 10^{-3}$, $L_{6}=(0.07\pm 0.08)\ 10^{-3}$ and
$L_{8}= (0.78\pm 0.18)\ 10^{-3}$
but we take these new values only as indicative because of their model dependence.

The phenomenological estimates \cite{Ecker:1988te,GomezNicola:2001as,Amoros:2000mc}
have been consistently computed at the same mass scale $\mu=M_\rho$.
Even if they show some dispersion, they are not far from the L$\sigma$M predictions in
Eq.~(\ref{predLisLsM}).
An improvement on the phenomenological values of these low energy constants could fix the mass of
the scalar singlet $M_{S_0}$ and the mean mass of the octet $M_{S_8}$.
For the latter, the well known estimate $L_{5}=(1.4\pm 0.5)\ 10^{-3}$ of Ref.~\cite{Ecker:1988te}
and Eq.~(\ref{LiSU3limit}) imply $M_{S_8}=(1.2\pm 0.2)$ GeV.
A confirmation of this central value for $L_{5}$ and a reduction of its error bars would represent
a success for the L$\sigma$M.
On the other hand a crucial test would be the sign of $L_{4}$,
which cannot be unambiguously fixed from the analyses in Ref.~\cite{Ecker:1988te},
that quotes $L_{4}=(-0.3\pm 0.5)\ 10^{-3}$.
A negative $L_{4}$ would imply via Eq.~(\ref{LiSU3limit}) $M_{S_0}>M_{S_8}$ representing a
serious problem for the L$\sigma$M which prefers $M_{S_0}<M_{S_8}$,
as indicated in Eqs.~(\ref{predmass8LsM}) and (\ref{predmass0LsM}).
Similarly, a large positive value for $L_{4}$ would represent a problem for part of our treatment:
one would then have $M_{S_0} \simeq m_K$ and poor convergence in some of our series expansions.
The L$\sigma$M prediction $2L_{6}+L_{8}=(0.6\pm 0.2)\ 10^{-3}$ in Eq.~(\ref{LiSU3limit})
fits perfectly with the central value for $2L_{6}+L_{8}$ in Eq.~(\ref{LisEcker2}) coming from
Ref.~\cite{Ecker:1988te} but the error bars are too large to draw a definite conclusion.
Something similar happens to the L$\sigma$M predictions for $L_{1,3}^S$ which are only a fraction
of the measurable $L_{1,3}$ and whose analysis is outside the scope of this work.

In conclusion, we have  compared the L$\sigma$M and ChPT predictions for pion and kaon dynamics.
The leading terms (order $p^2$ in the chiral expansion) are entirely reproduced by the L$\sigma$M,
as expected from chiral symmetry.
The next-to-leading terms (order $p^4$) are also consistently reproduced if one restricts to the
terms generated by scalar resonances.
In this case, the scalar contributions to the low energy constants of the (order $p^4$) ChPT Lagrangian
are fixed in terms of octet and singlet scalar masses, $M_{S_8}$ and $M_{S_0}$.
The corresponding expressions improve older results, which are recovered in the appropriate limits.
Precise values for the low energy constants should be useful to confirm or falsify the L$\sigma$M dynamics,
to fix some scalar resonance parameters and, hopefully, to shed some light on the
controversial nature of the lowest lying scalar states.

\begin{acknowledgments}
Work partly supported by the EURIDICE network (HPRN-CT-2002-00311),
and the Ministerio de Ciencia y Tecnolog\'{\i}a and FEDER, FPA2002-00748.
J.~L.~L.~M. acknowledges support from CONACyT under grant 37234-E.
\end{acknowledgments}

\end{document}